\documentclass[epj]{svjour}
\usepackage{graphics}

\newcommand{\QandA}[1]{%
\centerline{--~\parbox[t]{0.8\columnwidth}{\raggedright{}#1}}}

\newcommand{\eqref}[1]{(\ref{#1})}
\newcommand{\fnref}[1]{footnote \ref{fn:#1}}
\newcommand{\foreign}[1]{\textsl{#1}}
\newcommand{\pol}[1]{\mbox{\scriptsize\textsf{#1}}}  
\newcommand{\spol}[1]{\mbox{\tiny\textsf{#1}}}  
\newcommand{\half}{\frac{1}{2}}
\newcommand{\ket}[1]{{\left|#1\right\rangle}}
\newcommand{\bra}[1]{{\left\langle{#1}\right|}}
\newcommand{\braket}[3][]{{\left\langle{#2}{#1|}{#3}\right\rangle}}
\newcommand{\ketbra}[2]{\ket{#1}\bra{#2}}
\newcommand{\repr}{\mathrel{\widehat{=}}}
\newcommand{\column}[2][c]{{\left(\begin{array}{#1}#2\end{array}\right)}}
\newcommand{\I}{\mathrm{i}}\newcommand{\D}{\mathrm{d}}
\newcommand{\tr}[2][]{\mathrm{tr}_{#1}{\left\{#2\right\}}}
\newcommand{\determ}[1]{\det{\left\{#1\right\}}}
\newcommand{\WITH}{\quad\textrm{with}\quad}%
\newcommand{\AND}{\quad\textrm{and}\quad}%
\newcommand{\FOR}{\quad\textrm{for}\quad}%
\newcommand{\der}[2]{\frac{\D #1}{\D #2}}%
\newcommand{\pder}[2]{\frac{\partial #1}{\partial #2}}%
\newcommand{\sds}[2][\depth]{\raisebox{0pt}[\height][#1]%
{\footnotesize$\displaystyle{}#2$}}
\newcommand{\adj}{^{\dagger}}\newcommand{\phadj}{^{\vphantom{\dagger}}}
\newcommand{\Exp}[1]{\,\mathrm{e}^{\mbox{\footnotesize$#1$}}}
\newcommand{\power}[1]{^{\mbox{\footnotesize$#1$}}}
\newcommand{\rewop}[1]{_{\mbox{\footnotesize$#1$}}}
\newcommand{\phstar}{^{\vphantom{*}}}
\newcommand{\Dstate}[1]{\mbox{\small{#1}}}

\makeatletter
\newcommand{\magn}[1]{%
{\mathchoice{\magn@D{#1}}{\magn@T{#1}}{\magn@S{#1}}{\magn@SS{#1}}}}
\newlength{\@xx}\newlength{\@yy}\newlength{\@zz}
\newcommand{\magn@D}[1]{%
\settoheight{\@xx}{$\displaystyle\left|#1\right|$}%
\settodepth{\@yy}{$\displaystyle\left|#1\right|$}%
\addtolength{\@xx}{\@yy}%
{\,\rule[-\@yy]{\@zz}{\@xx}\,#1\,\rule[-\@yy]{\@zz}{\@xx}\,}}%
\newcommand{\magn@T}[1]{%
\settoheight{\@xx}{$\textstyle\left|#1\right|$}%
\settodepth{\@yy}{$\textstyle\left|#1\right|$}%
\addtolength{\@xx}{\@yy}%
{\,\rule[-\@yy]{\@zz}{\@xx}\,#1\,\rule[-\@yy]{\@zz}{\@xx}\,}}%
\newcommand{\magn@S}[1]{%
\settoheight{\@xx}{$\scriptstyle\left|#1\right|$}%
\settodepth{\@yy}{$\scriptstyle\left|#1\right|$}%
\addtolength{\@xx}{\@yy}%
{\,\rule[-\@yy]{\@zz}{\@xx}\,#1\,\rule[-\@yy]{\@zz}{\@xx}\,}}%
\newcommand{\magn@SS}[1]{%
\settoheight{\@xx}{$\scriptscriptstyle\left|#1\right|$}%
\settodepth{\@yy}{$\scriptscriptstyle\left|#1\right|$}%
\addtolength{\@xx}{\@yy}%
{\,\rule[-\@yy]{\@zz}{\@xx}\,#1\,\rule[-\@yy]{\@zz}{\@xx}\,}}%
\newcommand{\lmagn}{%
\settoheight{\@xx}{$\displaystyle\bigl( \bigr)$}%
\settodepth{\@yy}{$\displaystyle\bigl( \bigr)$}%
\addtolength{\@xx}{\@yy}%
{\,\rule[-\@yy]{\@zz}{\@xx}\,}}
\newcommand{\Lmagn}{%
\settoheight{\@xx}{$\displaystyle\Bigl( \Bigr)$}%
\settodepth{\@yy}{$\displaystyle\Bigl( \Bigr)$}%
\addtolength{\@xx}{\@yy}%
{\,\rule[-\@yy]{\@zz}{\@xx}\,}}%
\makeatother
\newcommand{\olmagn}[1]{\magn{\raisebox{0pt}[0pt][0pt]{$#1$}}}

\begin{document}


\title{On Quantum Theory\thanks{In sincere gratitude for many instructive
    discussions, I dedicate this essay to Professor Rudolf Haag on the
    occasion of his 90th birthday.}} 

\author{Berthold-Georg Englert}

\institute{Centre for Quantum Technologies and Department of Physics, National
University of Singapore, Singapore\newline \email{cqtebg@nus.edu.sg}}

\date{Posted on the arXiv on 24 August 2013}

\abstract{%
Quantum theory is a well-defined local theory with a clear interpretation.
No ``measurement problem'' or any other foundational matters are waiting to be
settled. 
} 

\maketitle

Quantum theory had essentially taken its final shape by the end of
the 1920s and, in the more than eighty years since then, has been spectacularly
successful and reliable --- there is no experimental fact, not a single one,
that contradicts a quantum-theoretical prediction.
Yet, there is a steady stream of publications
that are motivated by alleged fundamental problems:
We are told that quantum theory is ill-defined, that its interpretation is
unclear, that it is nonlocal, that there is an unresolved ``measurement
problem,''  and so forth.%
\footnote{\label{fn:NFP}%
  No references are given here or later; why point a finger at a particular 
  representative of one or the other community? 
  The pertinent essays in the 2009 \textit{Compendium of Quantum Physics} 
  \cite{Greenberger+2.ed:2009} cite the relevant papers.
  This no-finger-pointing policy applies throughout the chapter.
  In particular, we note that the quotes in \eqref{eq:0-Q1},
  \eqref{eq:0-Q2}, \eqref{eq:0-Q4}, and \eqref{eq:0-Q5} below
  are not made up.
  They are actual quotes from existing sources. 
  The sources are irrelevant, however, the quotes are 
  representatives of many similar statements.}

It may, therefore, be worth reviewing what quantum theory is and what it is
about.
While it is neither possible nor desirable to cover the ground exhaustingly,
we shall deal with the central issues and answer questions such as these:
\begin{quote}
\QandA{Is quantum theory well defined?}
\QandA{Is the interpretation of quantum theory clear?}
\QandA{Is quantum theory local?}
\QandA{Is quantum evolution reversible?}
\QandA{Do wave functions collapse?}
\QandA{Is there instant action at a distance?}
\QandA{Where is Heisenberg's cut?}
\QandA{Is Schr\"odinger's cat half dead and half alive?}
\QandA{Is there a ``measurement problem''?}
\end{quote}


\section{Physical theories: %
Phenomena, formalism, interpretation; preexisting concepts}
\label{sec:PFI}
In physics, a \emph{theory} has three defining constituents:
the physical phenomena, the mathematical formalism, and the interpretation.
The \emph{phenomena} are the empirical evidence about physical objects
gathered by passive observation, typical for astronomy and meteorology, or by
active experimentation in the laboratory.
The \emph{formalism} provides the adequate mathematical description of the
phenomena and enables the physicist to make precise quantitative predictions
about the results of future experiments.
The \emph{interpretation} is the link between the formalism and the phenomena.

Each physical theory relies on preexisting concepts, which reflect elementary
experiences and without which the theory could not be stated.
In Newton's classical mechanics \cite{Newton:87}, these concepts include mass
and force, and the theory deals with the motion of massive bodies under the
influence of forces.
Classical mechanics cannot provide any insight why there are masses and
forces.

Likewise, electric charge is a preexisting concept in Maxwell's
electromagnetism \cite{Maxwell:73}, which accounts for the forces exerted by
the electromagnetic field on charged bodies and how, in turn, the charges and
their currents give rise to the electromagnetic field.
Maxwell's theory cannot tell us why there are electricity and magnetism or why
there are charges and fields, nor is this the purpose of the theory.

\hspace*{-2.64pt}
One preexisting concept of quantum theory is the event, such as the emission
of a photon by an atom, the radioactive decay of a nucleus, or the ionization
of a molecule in a bubble chamber.
The formalism of quantum theory has the power to predict the probabilities
that the events occur, whereby Born's rule \cite{Born:26} is the link
between formalism and phenomenon.
But an answer to the question \emph{Why are there events?} cannot be given by
quantum theory.   

In addition, there are preexisting concepts that are common to all physical
theories: 
the three-dimensional space in which the natural phenomena happen, and the
time whose flow distinguishes the past from the future.
We must also acknowledge a fundamental philosophical concept of science ---
the conviction that it is possible, with means accessible to the human mind,
to give a systematic account of the natural phenomena.

\section{Events}\label{sec:events}
The formalism of quantum theory can be used to predict probabilities ---
probabilities for \emph{events}.
In the classic Stern--Gerlach experiment \cite{Gerlach+1:22},%
\footnote{The accounts by Friedrich and Herschbach \cite{Friedrich+1:03} and
  by Bernstein \cite{Bernstein:10} of this experiment's history are
  recommended reading.}
the silver atom eventually hits the glass plate, and we can calculate the
probability that the next atom will land in a certain region.
The hitting of the glass plate is localized in space and time and has an
element of irreversibility, \foreign{videlicet} the emission of
long-wavelength photons during the violent deceleration of the atom, which
photons are irretrievably lost.
Another example of an event is the radioactive $\alpha$-decay of a nucleus,
which is invariably accompanied by the bremsstrahlung associated with the
escaping $\alpha$ particle and the two now-superfluous electrons that are left
behind. 
Yet another example is the absorption of a photon by a semiconductor detector.
These examples illustrate the defining features of an event.

\textbf{First}, an event is well localized in space and time.
Haag \cite{Haag:90} emphasized, and rightly so, that short-range interactions
are crucial for the localization of an event and that events are linked by
particles, so that a causal history evolves --- think, for instance, of the
sequence of ionization events in a bubble chamber.
By their spatial-temporal relations, the events give meaning to the space-time
structure, the great stage for all natural phenomena.
 
\textbf{Second}, an event is irreversible, it leaves a document behind, a
definite trace. 
It has often been emphasized that an amplification is necessary to bring the
event to the attention of the experimenter and that the amplification is an
irreversible process; this is true, but it downplays (or even ignores) the
irreversibility of the event itself, which is independent of human
observation.
If you don't have an event in the first place, there is nothing you can
amplify.
In the bubble chamber situation above, the amplification turns the sequence of
ionization events into a string of bubbles that reveal the track of the
charged particle to the human eye.

\textbf{Third}, events are randomly realized.
Before an event occurs, there are usually several, perhaps many, different
events that could happen, each of them with a certain probability of
occurrence.
Eventually, a definite one of the possible events will be realized.
We cannot predict which one will be the case --- because this is unknowable,
not because we are missing a crucial piece of information.
 
Quantum theory can be very helpful in understanding quite a lot of the details
of, say, the ionization of a molecule by a charge that passes by.
We can calculate, with reasonable accuracy, the probability of ionization, the
probability that the molecule gets energetically excited without being ionized
and subsequently emits one or more photons, and the probability that the
molecule remains unaffected by the charged particle.
Whether the ionization event happens or not, we cannot predict --- the
\emph{principle of random realization} (Haag in \cite{Haag:13}) ensures that
the events do happen in accordance with their probabilities of occurrence.

Born's rule tells us these probabilities --- the probability
for the ionization event, the probability for the photon-emission event, and
the no-event probability for the molecule remaining unaffected.
For that to have any meaning, the existence of events must be accepted in the
first place.
In this sense, then, the event is a preexisting concept of quantum theory.
We cannot formulate the theory without this concept.

\section{Measurements, Born's rule}\label{sec:MeasBorn}
\subsection{An example}\label{sec:MeasBorn-1}
\begin{figure}[t]
\centerline{\includegraphics{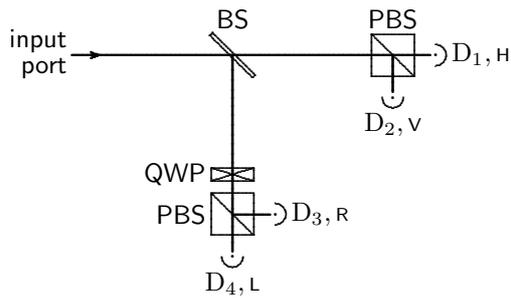}}  
\caption{\label{fig:4POM}%
A four-outcome measurement of photon polarization.
An incident photon has equal chance of transmission or reflection at the beam
splitter (\textsf{BS}). 
The polarizing beam splitters (\textsf{PBS}) reflect vertically polarized
photons and transmit horizontally polarized photons.
Photons transmitted at the \textsf{BS} are probed for  horizontal (\pol{H})
and vertical (\pol{V}) polarization and detected by detectors $\mathrm{D}_1$
and $\mathrm{D}_2$, respectively. 
The combination of quarter-wave plate (\textsf{QWP}) and \textsf{PBS} probes
the photons reflected at the \textsf{BS} for right-handed (\pol{R}) or
left-handed (\pol{L}) circular polarization, with respective detectors
$\mathrm{D}_3$ and $\mathrm{D}_4$. 
}
\end{figure}

Consider the set-up sketched in Fig.~\ref{fig:4POM} where photons are incident
on a half-transparent mirror that serves as a symmetric beam splitter
(BS). Whether transmitted or reflected, the photons then encounter a
polarizing beam splitter (PBS) that reflects vertically (\pol{V})
polarized photons and transmits horizontally (\pol{H}) polarized photons.
For simplicity, we assume that the BS is symmetric for all polarizations, so
that the next photon to arrive has an equal chance of being reflected or
transmitted at the BS, irrespective of its polarization.
The photons that are reflected pass through a quarter-wave plate (QWP), which
converts right-handed (\pol{R}) circular polarization to vertical
polarization and left-handed (\pol{L}) polarization to horizontal
polarization. 
The four detectors $\mathrm{D}_1$, \dots, $\mathrm{D}_4$ in the output ports
of the PBSs, therefore, distinguish between \pol{H} and \pol{V} polarization
or between \pol{R} and \pol{L} polarization of the incoming photon.

The construction of the single-photon detector for a particular outcome is
largely irrelevant, except that its efficiency depends on which physical process
is used for absorbing the photon and how the absorption event is amplified to
finally make the detector ``click.''
With photons arriving one by one, separated in time by more than the dead-time
of the detectors, the measurement result consists of a sequence of detector
clicks. 
Quantum theory cannot predict which sequence will be recorded for the next,
say, one hundred photons but, for the next photon to arrive, it tells us
reliably the probability of clicking for each detector, provided we know the
polarization of this incoming photon.

Which one of the four detectors will be triggered by the next photon, which
one of the four possible events will be the actual one, this is not just
unknown, it is unknowable.
One of the events will be randomly realized.

That quantum theory cannot predict which detector will click for the next
photon, does not imply that quantum theory is incomplete.
There is nothing missing; it is not possible to add further elements to the
quantum-theoretical formalism or its interpretation and then have the power of
making such predictions.
There are no consistent modifications of quantum theory that will achieve this
without getting wrong predictions in other situations.
The question \textit{Can Quantum-Mechanical Description of Physical Reality be
  Considered Complete?}  
that Einstein, Podolsky, and Rosen asked in 1935 \cite{Einstein+2:35}
has a clear answer: Yes, quantum theory gives a complete description of the
phenomena.
Bohr's reply \cite{Bohr:35a,Bohr:35b} hit the mark.

In the experiment of Fig.~\ref{fig:4POM}, it is in fact impossible to prepare
the photon such that a particular detector will certainly click.
Yes, we can make sure that one of the detectors will not click ---
\pol{H}-polarized photons will never give rise to a click of $\mathrm{D}_2$, 
for instance, but then the photon has probabilities of $\half$,
$\frac{1}{4}$, $\frac{1}{4}$ of entering one of the other three detectors.
Of course, what we have here is just an example of the fundamental
probabilistic nature of quantum processes --- an illustration of the
principle of random realization at work.

It should be noted that the reflection or transmission of the photon at the BS
does not constitute an event.
The photon needs a short-range interaction with a partner to give rise to an
event. 
If no such partner is available, we could complete a 
Mach--Zehnder
interferometer by adding mirrors and another BS, after which it would be
unknowable if the exiting photon was reflected or transmitted at the BSs.

\subsection{Mathematical description}\label{sec:MeasBorn-2}
We represent the kets for the polarization state of a photon by two-component
columns, for which we use these conventions:
\begin{eqnarray}\label{eq:0-1}
 && \ket{\pol{V}}\repr\column{1\\0}\,,\quad
  \ket{\pol{H}}\repr\column{0\\1}\,,\nonumber\\
 &&\ket{\pol{R}}\repr\frac{1}{\sqrt{2}}\column{1\\\I}\,,\quad
  \ket{\pol{L}}\repr\frac{1}{\sqrt{2}}\column{\I\\1}\,.
\end{eqnarray}
Accordingly, the statistical operator for the polarization degree of freedom
for the incoming photon is represented by a $2\times2$ matrix,
\begin{equation}
  \label{eq:0-2}
  \rho=\column[cc]{\ket{\pol{V}}&\ket{\pol{H}}}
       \column[cc]{\rho_{\spol{VV}}^{\ } & \rho_{\spol{VH}}^{\ } \\
                   \rho_{\spol{HV}}^{\ } & \rho_{\spol{HH}}^{\ }}
       \column{\bra{\pol{V}}\\\bra{\pol{H}}}
     \repr\column[cc]{\rho_{\spol{VV}}^{\ } & \rho_{\spol{VH}}^{\ }\\
                   \rho_{\spol{HV}}^{\ } & \rho_{\spol{HH}}^{\ }}\,,
\end{equation}
and the probability that detector $\mathrm{D}_1$ will click for the next
photon is
\begin{equation}
  \label{eq:0-3}
  p_1=\half\eta_1\rho_{\spol{HH}}^{\ }
     =\half\eta_1\bra{\pol{H}}\rho\ket{\pol{H}}=\tr{\rho\Pi_1}
\end{equation}
with the probability operator
\begin{equation}
  \label{eq:0-4}
  \Pi_1=\ket{\pol{H}}\half\eta_1\bra{\pol{H}}
        \repr\half\eta_1\column[cc]{0&0\\0&1}\,.
\end{equation}
Here, the factor of $\half$ is the transmission probability of the BS and
$\eta_1$ is the detection efficiency of $\mathrm{D}_1$, the conditional
probability that a click is triggered by an incident photon.

Similarly, the probability operators for the other three detectors are%
\footnote{For simplicity, we ignore here and in \eqref{eq:0-3} the possibility
  of ``dark counts'' --- spontaneous detector clicks not triggered by an
  incident photon.}
\begin{eqnarray}\label{eq:0-5}
  \Pi_2&=&\ket{\pol{V}}\half\eta_2\bra{\pol{V}}
        \repr\half\eta_2\column[cc]{1&0\\0&0}\,,
\nonumber\\[1ex]
  \Pi_3&=&\ket{\pol{R}}\half\eta_3\bra{\pol{R}}
        \repr\frac{1}{4}\eta_3\column[cc]{1&-\I\\ \I&\ 1}\,,\nonumber\\[1ex]
  \Pi_4&=&\ket{\pol{L}}\half\eta_4\bra{\pol{L}}
        \repr\frac{1}{4}\eta_4\column[cc]{\ 1& \ \I\\-\I&\ 1}\,.  
\end{eqnarray}
These need to be supplemented by the probability operator $\Pi_0$ for the
possibility that the photon escapes detection (no click),
\begin{equation}\label{eq:0-6}
  \Pi_0=1-(\Pi_1+\Pi_2+\Pi_3+\Pi_4)\,,
\end{equation}
to make up the probability-operator measurement (POM%
\footnote{POM, with its emphasis on ``probability'' and ``measurement,'' is
preferable quantum physics jargon.
The mathematical term POVM
(positive-operator-valued measure) makes reference to measure theory.
Note that ``theory'' here identifies a mathematical discipline and does not
have the meaning of Sec.~\ref{sec:PFI}. }) for the set-up of
Fig.~\ref{fig:4POM}.

With the probability operators $\Pi_k$ at hand, the respective probabilities
$p_k$ are obtained by \emph{Born's rule},
\begin{equation}\label{eq:0-7}
  p_k=\tr{\rho\Pi_k}\,.
\end{equation}
As already noted in Sec.~\ref{sec:PFI} above,
Born's rule is a defining element of the interpretation of quantum theory, it
links symbols of the mathematical formalism with the phenomenology: the
statistical operator, which summarizes what we know about the
polarization state of the incoming photon, and the probability operator, which
represents the apparatus, with the probability of a detector click, which we
can determine in an experiment by observing sufficiently many repetitions of
the situation ``single incident photon triggers a detector click.''

\subsection{Generic measurements}\label{sec:GenMeas}
\begin{figure}[t]
\centerline{\includegraphics{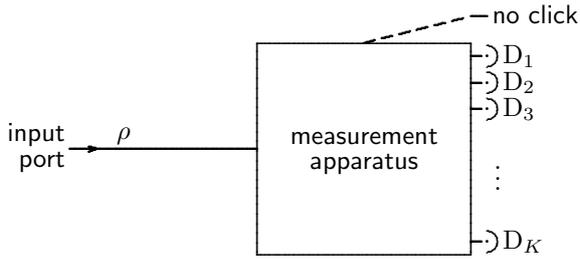}} 
\caption{\label{fig:genPOM}%
A generic measurement.
A physical object (photon, electron, atom, molecule, \dots), whose relevant
quantum degrees of freedom are described by the statistical operator $\rho$,
enters the measurement apparatus and triggers a click of one of the $K$
detectors $\mathrm{D}_1$, $\mathrm{D}_2$, \dots, $\mathrm{D}_K$, 
or it escapes detection (no click).
}
\end{figure}

The particular example of Fig.~\ref{fig:4POM} illustrates the general
measurement scenario.
Let us briefly recall the generic properties of a measurement in the quantum
realm, as depicted in Fig.~\ref{fig:genPOM}.
The relevant degrees of freedom of the quantum system are described by the
statistical operator $\rho$, and the $K$ outcomes of the POM have
probability operators $\Pi_1$, $\Pi_2$, \dots, $\Pi_K$.

The statistical operator is positive and normalized to unit trace,
\begin{equation}
  \label{eq:0-8}
  \rho\geq0\,,\quad\tr{\rho}=1\,,
\end{equation}
and the probability operators are positive and have unit sum,
\begin{equation}
  \label{eq:0-9}
  \Pi_k\geq0\,,\quad\sum_{k=0}^K\Pi_k=1\,,
\end{equation}
where we include $\Pi_0$, the probability operator for the ``no click'' case.
The probability that the $k$th detector clicks is given by 
Born's rule
\eqref{eq:0-7}, which also applies for the no-click probability $p_0$.
As a basic check of consistency, one can verify without effort that the
probabilities are positive and have unit sum,
\begin{equation}
  \label{eq:0-10}
  p_k\geq0\,,\quad\sum_{k=0}^Kp_k=1\,.
\end{equation}

The probabilities $p_k$ are conditional probabilities --- they are conditioned
on what we know about the physical system, which information is encoded in the
statistical operator that we use for our probabilistic predictions.
Someone else may have different knowledge about the system and, therefore,
use a different statistical operator and arrive at different values of the
probabilities $p_k$. 
Yet, her predictions and ours, though different, will both be correct and
verifiable.

\section{Statistical operators}
\subsection{Mixtures, blends, and as-if realities}\label{sec:as-if}
The general form of a statistical operator for the polarization degree of
freedom of a photon is given in \eqref{eq:0-2}, where the matrix elements are
restricted by
\begin{eqnarray}\label{eq:0-10a}
&&\rho_{\spol{AB}}^{\ }={\rho_{\spol{BA}}^{\ }}^* 
\FOR \pol{A},\pol{B}=\pol{H},\pol{V}\,,\nonumber\\
&&\rho_{\spol{VV}}^{\ }+\rho_{\spol{HH}}^{\ }=1\,,\AND
\rho_{\spol{VV}}^{\ }\rho_{\spol{HH}}^{\ }\geq\rho_{\spol{VH}}^{\ }\rho_{\spol{HV}}^{\ }\,.
\end{eqnarray}
Let us consider a $\rho$ with a positive nonmaximal determinant,
\begin{equation}\label{eq:0-10b}
  \frac{1}{4}>\determ{\rho}=
\rho_{\spol{VV}}^{\ }\rho_{\spol{HH}}^{\ }-\rho_{\spol{VH}}^{\ }\rho_{\spol{HV}}^{\ }>0\,,
\end{equation}
so that the polarization state is impure (${\rho^2\neq\rho}$) 
and $\rho$ has the two positive eigenvalues
\begin{equation}\label{eq:0-10c}
  \left.\begin{array}{l}r_1\\r_2\end{array}\right\}=
  \half\pm\half\sqrt{1-4\determ{\rho}\,}\,,\quad
  1>r_1>r_2>0\,.
\end{equation}
The rank-one projectors to the respective eigenspaces are
\begin{equation}\label{eq:0-10d}
  \ketbra{r_1}{r_1}=\frac{\rho-r_2}{r_1-r_2}\AND  
  \ketbra{r_2}{r_2}=\frac{r_1-\rho}{r_1-r_2}\,,
\end{equation}
and
\begin{equation}\label{eq:0-10e}
  \rho=\ket{r_1}r_1\bra{r_1}+\ket{r_2}r_2\bra{r_2}
\end{equation}
is the spectral decomposition of $\rho$.

If $\rho$ refers to a large ensemble of photons, we could say that it is
\emph{as if} $\rho$ came about by having fractions $r_1$ and $r_2$ of photons
in the polarization states associated with kets $\ket{r_1}$ and $\ket{r_2}$,
respectively. 
Speaking of the one photon of interest, regarded as the representative of a
Gibbs ensemble of photons, it is \emph{as if} that photon
had respective chances $r_1$ and $r_2$ of being of the $\ket{r_1}$ or
$\ket{r_2}$ kind. 
This is but one of very many \emph{as-if realities} for the one statistical
operator $\rho$.
For example, we can also write
\begin{equation}\label{eq:0-10f}
  \rho=\ket{r_+}\half\bra{r_+}+\ket{r_-}\half\bra{r_-}
\end{equation}
with
\begin{equation}
  \label{eq:0-10g}
  \ket{r_{\pm}}=\ket{r_1}\sqrt{r_1}\pm\ket{r_2}\sqrt{r_2}\,.
\end{equation}
Now it is as if the photon had equal chances of being in the polarization
states for kets $\ket{r_+}$ and $\ket{r_-}$.
Hereby, each choice for the relative phase between $\ket{r_1}$ and $\ket{r_2}$,
when factoring the projectors in \eqref{eq:0-10d} into ket-bra products, gives
different $r_{\pm}$ kets in \eqref{eq:0-10g} and, therefore, different
projectors on the right-hand side of \eqref{eq:0-10f} and a different as-if
reality to go with them.
Any convex sum of \eqref{eq:0-10e} and \eqref{eq:0-10f} identifies yet another
as-if reality.

We have the same mixed state $\rho$, the same \emph{mixture}, and different
\emph{blends} for it,%
\footnote{We are adopting S\"u\ss{}\-mann's fitting terminology
who distinguishes between \foreign{Gemisch} (mixture) and
\foreign{Gemenge} (blend) \cite{Sussmann:58}.}
each of them offering a different as-if reality. 
All that matters is the mixture, not how it is blended as a convex sum of pure
states.  
Born's rule \eqref{eq:0-7} makes no reference to the blend and
its as-if reality, it only cares about the mixture $\rho$.
There is no way of distinguishing between the different blends of
\eqref{eq:0-10e} and \eqref{eq:0-10f} or of their many convex sums;
no experiment can tell whether this as-if reality or that one is ``what really
is the case.'' Attempts at discriminating
between blends founder on fundamental principles, such as 
Einsteinian causality \cite{Gisin:89,Gisin:90}.
All ways of blending a mixture are equally good, all as-if realities are
equally virtual.
In particular, the spectral decomposition in \eqref{eq:0-10e} and the as-if
reality associated with it have no special significance.

Of course, these observations about mixtures, blends, and as-if realities 
--- here illustrated in the simple context of photon polarization --- 
carry over to more general situations.
Whenever the statistical operator is not a rank-one projector, there will be a
plethora of blends for that mixture and a corresponding abundance of as-if
realities.

\subsection{Purification}\label{sec:purify}
A mixture that is blended from $J$ rank-one ingredients,
\begin{equation}\label{eq:0-10h}
  \rho=\sum_{j=1}^J\ket{a_j}w_j\bra{a_j}
      =\sum_{j,j'=1}^J\ket{a_j}\sqrt{w_j}\,\delta_{j,j'}\,\sqrt{w_{j'}}\bra{a_{j'}}
\end{equation}
with positive weights of unit sum ($w_j>0$, $\sum_jw_j=1$), can be written as
the partial trace of a rank-one projector of a bipartite system,
\begin{equation}\label{eq:0-10i}
  \rho=\tr[2]{\vphantom{\big|}\ketbra{\ }{\ }}
\end{equation}
with
\begin{equation}\label{eq:0-10j}
  \ket{\ }=\sum_{j=1}^J\ket{a_j\,b_j}\sqrt{w_j}\,,
\end{equation}
where the $b$ kets are orthonormal, $\braket{b_j}{b_{j'}}=\delta_{j,j'}$.
Since the blend that we start with in \eqref{eq:0-10h} is not unique and any
orthonormal set of $b$ kets of any auxiliary second system will serve the
purpose, there are many different ket-bras $\ketbra{\ }{\ }$ that one can use
in \eqref{eq:0-10i} equally well. 

In particular situations, such as the one discussed in
Sec.~\ref{sec:notspooky} below, we are really dealing with a bipartite system
in a pure state (or nearly so), and then a physically meaningful ket 
$\ket{\ }$ may be available. 
In the generic case, however, we have our statistical operator $\rho$ for the
physical system of interest and the \emph{purification} of \eqref{eq:0-10i}
with \eqref{eq:0-10j} is just a mathematical procedure that might be helpful
in performing a certain calculation but has no physical significance
whatsoever.

Yet, the ``church of the larger Hilbert space''
offers membership to those who believe that there is always such a larger
physical system in a pure state.
Now, it is already quite challenging to acquire sufficient information 
about a few degrees of freedom to justify, as a good approximation, 
the use of a rank-one projector as the statistical operator. 
How, then, can we have a pure state for the unknown degrees of freedom of a
fictitious larger system that exists only in our fantasy?  
Of course, we cannot. 

It appears that sloppy language is one source of this misconception.
One often speaks of ``the wave function of the electron'' or
``the statistical operator of the photon'' rather than ``our wave
function for the electron'' or ``our statistical operator for the photon.''
The precise ``our~\dots~for'' phrases leave no doubt that we are describing
the electron by this wave function or the photon by that statistical operator, 
whereas the sloppy ``the~\dots~of'' wordings wrongly suggest an independent
existence of the wave function or the statistical operator.
When this suggestion is taken seriously, the church gates stand wide open.

Is there not the one largest system, the universe, with the physical system of
interest as a small part of it?
Yes, there is the universe, but no-one has a wave function \emph{for} the
universe, and there is no wave function \emph{of} the universe.

\section{Evolution}\label{sec:evolution}
\subsection{Equations of motion}\label{ssec:eqsmotion}
In Born's rule \eqref{eq:0-7}, both the statistical operator
$\rho$ and the probability operator $\Pi_k$ are functions of the dynamical
variables --- collectively symbolized by $Z(t)$ --- with, possibly, a
parametric time dependence as well. 
\emph{All} such operator-valued functions evolve in accordance with
\emph{Hei\-sen\-berg's equation of motion},
\begin{equation}
  \label{eq:0-11}
  \der{}{t}f\bigl(Z(t),t\bigr)=\pder{}{t}f\bigl(Z(t),t\bigr)
+\frac{1}{\I\hbar}\bigl[f\bigl(Z(t),t\bigr),H\bigl(Z(t),t\bigr)\bigr]\,,
\end{equation}
the quantum analog of Hamilton's equation of motion
for phase-space functions.
In \eqref{eq:0-11}, 
$H\bigl(Z(t),t\bigr)$ is the Hamilton operator, 
itself a function of
this kind, and the commutator term accounts for the dynamical time dependence
of $f\bigl(Z(t),t\bigr)$.
Whereas $\Pi_k$ and $H$ may or may not have a parametric time dependence, the
parametric time dependence of the statistical operator compensates for its
dynamical time dependence because $\rho$ is constant,
\begin{eqnarray}
  \label{eq:0-12}
  \der{}{t}\rho\bigl(Z(t),t\bigr)&=&0\,,\nonumber\\
  \pder{}{t}\rho\bigl(Z(t),t\bigr)&=&
   \frac{1}{\I\hbar}\bigl[H\bigl(Z(t),t\bigr),\rho\bigl(Z(t),t\bigr)\bigr]\,.
\end{eqnarray}
The latter equation, really a special case of \eqref{eq:0-11}, is known as 
\emph{von Neumann's equation of motion},
the quantum analog of Liouville's equation of motion
for phase-space densities.

Since the von Neumann equation concerns solely the parametric time dependence
of $\rho$, it does not matter to which common time $t_0$ the dynamical
operators refer, 
\begin{equation}\label{eq:0-13}
   \pder{}{t}\rho\bigl(Z(t_0),t\bigr)=
   \frac{1}{\I\hbar}\bigl[H\bigl(Z(t_0),t\bigr),
                          \rho\bigl(Z(t_0),t\bigr)\bigr]\,,
\end{equation}
and since a unitary transformation relates the dynamical operators at time $t$
to those at time $t_0$, we also have
\begin{eqnarray}
  \label{eq:0-14}
  p_k(t)&=&\tr{\Pi_k\bigl(Z(t),t\bigr)\rho\bigl(Z(t),t\bigr)}\nonumber\\
        &=&\tr{\Pi_k\bigl(Z(t_0),t\bigr)\rho\bigl(Z(t_0),t\bigr)}\,.
\end{eqnarray}
According to standard terminology, these two expressions for $p_k(t)$ 
are ``in the Heisenberg picture'' and ``in the Schr\"o\-din\-ger picture,'' 
respectively.  
The central von Neumann equation 
in \eqref{eq:0-12} or \eqref{eq:0-13}, however, is picture independent. 
The parametric time dependence of the statistical operator is what it is.

The product $\Pi_k\rho$ is also an operator that obeys
Heisenberg's equation of motion \eqref{eq:0-11}, and the
two versions of \eqref{eq:0-14} give
\begin{equation}\label{eq:0-14a}
  \der{}{t}p_k(t)=\tr{\der{}{t}(\Pi_k\rho)}=\tr{\pder{}{t}(\Pi_k\rho)}
\end{equation}
for the time derivative of the $k$th probability.
The difference of the two traces is the vanishing trace of the commutator of
$\Pi_k\rho$ with the Hamilton operator.
In this sense, then, the Heisenberg picture pays
attention to the full time dependence, whereas the 
Schr\"odinger picture cares only
about the parametric time dependence.
This is particularly well visible when the probability operator $\Pi_k$ has no
parametric time dependence (the same set-up is used at different times), so
that \eqref{eq:0-14a} becomes
\begin{equation}\label{eq:0-14b}
  \der{}{t}p_k(t)=\tr{\der{\Pi_k}{t}\rho}=\tr{\Pi_k\pder{\rho}{t}}
\end{equation}
after the product rule is applied,
\begin{equation}
  \label{eq:0-14c}
  \der{}{t}(\Pi_k\rho)=\der{\Pi_k}{t}\rho\,,\quad
\pder{}{t}(\Pi_k\rho)=\Pi_k\pder{\rho}{t}\,.
\end{equation}
The first of these statements is an identity that is generally valid, the
second holds only when $\sds{\pder{}{t}}\Pi_k=0$.

If we denote by $\bra{a,t}$ the common eigenbras of a maximal set $A(t)$ of
commuting dynamical variables, a subset of the collection $Z(t)$, with $a$
symbolizing the set of eigenvalues, $\bra{a,t}A(t)=a\bra{a,t}$,
then we have the
\emph{Schr\"odinger equation}
\begin{equation}
  \label{eq:0-15}
  \I\hbar\pder{}{t}\bra{a,t}=\bra{a,t}H\bigl(Z(t),t\bigr)
\end{equation}
for the relation between $\bra{a,t+\D t}$ and $\bra{a,t}$.
The matrix elements of a rank-one statistical operator, 
$\rho=\ketbra{\ }{\ }$, in this $a$ basis,  
\begin{equation}
  \label{eq:0-16}
  \bra{a,t}\rho\ket{a',t}=\psi(a,t)\psi(a',t)^*\,,
\end{equation}
are products of the probability amplitudes
\begin{equation}
  \label{eq:0-17}
  \psi(a,t)=\braket{a,t}{\ }
\end{equation}
and their complex conjugates.
The von Neumann equation \eqref{eq:0-12} is then equivalent to
\begin{equation}
  \label{eq:0-18}
  \I\hbar\pder{}{t}\psi(a,t)=\sum_{a'}\bra{a,t}H\bigl(Z(t),t\bigr)\ket{a',t}
                             \psi(a',t)
\end{equation}
or 
\begin{equation}
  \label{eq:0-19}  \I\hbar\pder{}{t}\Psi(t)=\mathcal{H}\Psi(t)\,,
\end{equation}
if we collect the probability amplitudes $\psi(a,t)$ in the column $\Psi(t)$,
the \emph{wave function}, and the matrix elements of $H$ in the matrix
$\mathcal{H}$. 
The corresponding matrix version of the von Neumann
equation itself is then
\begin{equation}
  \label{eq:0-18a}
  \pder{}{t}\varrho=\frac{1}{\I\hbar}\bigl[\mathcal{H},\varrho\bigr]\,,
\end{equation}
where $\varrho=\Psi\Psi\adj$ is the matrix that represents the statistical
operator $\rho=\ket{\ }\bra{\ }$. 
Of course, \eqref{eq:0-19} is the most familiar version of the
Schr\"odinger equation, 
as it applies to a wave function. 

Expressions such as \eqref{eq:0-16} may convey the false impression that
the statistical operator depends on time.
Yes, its matrix elements do depend on time, whereas we have
$\sds{\der{}{t}}\rho=0$ for the time dependence of $\rho$.
It is the time-dependent basis of bras that gives time dependence to the wave
function in \eqref{eq:0-17}, whereas the state ket $\ket{\ }$ is time
independent.
In
\begin{equation}\label{eq:0-18b}
  \ket{\ }=\sum_a\ket{a,t}\psi(a,t)=\sum_{a'}\ket{a',t'}\psi(a',t')\,,
\end{equation}
which holds for any time $t$ or $t'$, 
the basis kets $\ket{a,t}$ and $\ket{a',t'}$ as well as the probability 
amplitudes $\psi(a,t)$ and  $\psi(a',t')$ change in time, 
but $\ket{\ }$ does not.
This is, of course, as it should be because $\rho=\ket{\ }\bra{\ }$ represents
our knowledge about the preparation of the system and that makes reference to
some definite instant in the past when this knowledge was acquired.

In passing, we note that some textbook authors have failed to distinguish
clearly between the state ket $\ket{\ }$ (an abstract object) and its wave
function $\Psi$ (a basis-depen\-dent numerical representation of $\ket{\ }$),
between the Hamilton operator $H$ and its matrix $\mathcal{H}$, and between
the statistical operator $\rho$ and its matrix $\varrho$.
In their books, then, \eqref{eq:0-19} is written as
$\I\hbar\sds{\pder{}{t}}\ket{\Psi(t)}=H\ket{\Psi(t)}$, as if there were a
time-dependent state ket $\ket{\Psi(t)}$ that obeys a 
Schr\"odinger equation.
As a consequence, it then seems that we have a time-dependent statistical
operator $\rho=\ket{\Psi(t)}\bra{\Psi(t)}$ although, in fact, $\rho$ is
constant. 
The related failure of distinguishing between the total and the parametric
time dependence then gives rise to the misleading statement that equations
such as \eqref{eq:0-14b} demonstrate that 
``in the Schr\"odinger picture
$\rho$ depends on time and $\Pi_k$ is constant, whereas $\rho$ is constant
and $\Pi_k$ depends on time in the Heisenberg picture''
and the like.

\subsection{Time-reversal symmetry and irreversible evolution%
\protect\footnotemark}
\footnotetext{This section follows \cite{Englert:91/97} closely.}
\label{ssec:irreversible}
Let us consider the Schr\"odinger equation
for the position wave function of a particle of mass $M$ that moves along the
$x$ axis under the influence of the forces associated with the potential
energy $V(x)$,
\begin{equation}\label{eq:0-51}
  {\left(\frac{\partial}{\partial t}+\I\Omega\right)}\psi(x,t)=0\,,
\end{equation}
where the differential operator $\Omega$ is given by
\begin{equation}\label{eq:0-52}
  \hbar\Omega
=\frac{1}{2M}{\left(\frac{\hbar}{\I}\frac{\partial}{\partial x}\right)}^2
+V(x)>0\,.
\end{equation}
Here, for simplicity, the physical requirement that the energy is bounded from
below is equivalently replaced by insisting on the positivity of $\Omega$. 
The time-reversed wave function 
\begin{equation}\label{eq:0-53}
  (\Theta_T^{\ }\psi)(x,t)=\psi(x,2T-t)^*
\end{equation}
is also a solution of the 
Schr\"odinger equation
\eqref{eq:0-51}, 
\begin{equation}\label{eq:0-54}
    {\left(\frac{\partial}{\partial t}+\I\Omega\right)}\psi(x,2T-t)^*=0\,.
\end{equation}
This \emph{time-reversal symmetry} of the Schr\"odinger equation does not
imply, however, that quantum evolution is reversible.

One can imagine the undoing of all  
changes in $\psi(x,t)$ that have occurred
between, say, ${t=0}$ and ${t=T}$ by first applying $\Theta_T^{\ }$, followed
by evolution under the same dynamics for ${T<t<2T}$, and final application of 
$\Theta_{2T}^{\ }$, so that an instant $\tau$ after ${t=2T}$ one has the same
wave function as at time ${t=\tau}$, that is
\begin{eqnarray}\label{eq:0-55}
  \psi(x,2T+\tau)&=&\Exp{-\I\Omega\tau}\Theta_{2T}^{\ }
              \Exp{-\I\Omega T}\Theta_{T}^{\ }\Exp{-\I\Omega T}\psi(x,0)
\nonumber\\ &=& \psi(x,\tau)=\Exp{-\I\Omega\tau}\psi(x,0)
\end{eqnarray}
for all $x$.
But this cannot be achieved in any real sense.

The changes symbolized by $\Theta_{T}^{\ }$ and $\Theta_{2T}^{\ }$ must be of
dynamical origin themselves, which they cannot be because these operators are
not unitary; they are anti-unitary,
\begin{equation}\label{eq:0-56}
  \Theta_T^{\ }(\lambda_1\phstar\psi_1\phstar+\lambda_2\phstar\psi_2\phstar)
=\lambda_1^*\,\Theta_T^{\ }\psi_1\phstar+\lambda_2^*\,\Theta_T^{\ }\psi_2\phstar\,.
\end{equation}
It is true that all that matters is the product 
\begin{equation}\label{eq:0-57}
  \Theta_{2T}^{\ }\Exp{-\I\Omega T}\Theta_{T}^{\ }=\Exp{\I\Omega T}\,,
\end{equation}
and this is unitary.
But for most $\Omega$ the right-hand side is \emph{not} of the form
$\Exp{-\I\widetilde{\Omega}T}$ with a physical, positive $\widetilde{\Omega}$.
It follows that the undoing of all evolutionary changes is impossible, at
least if one just wants to exploit the time-reversal symmetry of the 
Schr\"odinger equation.

A more general scheme, not so tightly bound to elementary nonrelativistic
quantum mechanics, is what one could call the \emph{Humpty-Dumpty problem},
which name refers to the rhymed riddle (solution: egg) that
succinctly expresses folk wisdom about irreversibility.%
\footnote{You can find the riddle in \cite{Schwinger+2:88}.}
Here, then, is the 
\begin{equation}\label{eq:0-58}
  \parbox{0.8\columnwidth}{%
\centerline{\textbf{Humpty-Dumpty problem}}
Given a statistical operator $\rho$ and a Hamilton operator ${H_1(t)>0}$,
acting from ${t=0}$ to ${t=T}$, find $\widetilde{T}$ and ${H_2(t)>0}$, acting
from ${t=T}$ to ${t=\widetilde{T}}$, such that the unitary evolution operators
$U(T,0)$ and $U(\widetilde{T},T)$, constructed as time-ordered exponentials from
$H_1(t)$ and $H_2(t)$, respectively, give an over-all evolution operator
$U(\widetilde{T},0)= U(\widetilde{T},T)U(T,0)$ that has no effect on $\rho$: 
${U(\widetilde{T},0)\,\rho=\rho\, U(\widetilde{T},0)}$.\footnotemark}
\end{equation}
\footnotetext{More generally, one could also consider ancilla-assisted
  evolution between $T$ and $\widetilde{T}$, so that the net effect on $\rho$
  would come from a completely positive map that may not be describable by
  a single unitary operator $U(\widetilde{T},T)$. 
  The main conclusion that, as a rule, quantum evolution is irreversible is
  not affected, however.}%
A solution is found when we have
\begin{eqnarray}\label{eq:0-59}
U(\widetilde{T},T)\adj&=&
  \exp{\left(+\frac{\I}{\hbar}\int\limits\power{\widetilde{T}}\rewop{T}
\D t\,H_2(t)\right)}_<\nonumber\\&\mathrel{\dot{=}}& 
\exp{\left(-\frac{\I}{\hbar}\int\limits\power{T}\rewop{0}
\D t\,H_1(t)\right)}_>=U(T,0)\quad
\end{eqnarray}
with standard time ordering in the second line and the reverse order in the
first, and the dotted equal sign means ``same effect on $\rho\,$,'' that is
\begin{equation}\label{eq:0-59'}
  U(\widetilde{T},T)\adj\,\rho\,U(\widetilde{T},T)
 =U(T,0)\,\rho\,U(T,0)\adj\,.
\end{equation}
In view of the positivity of both $H_1(t)$ and $H_2(t)$ this can only be
realized under exceptional circumstances.
The question is then: For which $\rho$ and $H_1(t)$ is it possible at all?
There are, of course, two very important rules in this game, namely
\begin{equation}\label{eq:0-60}
\parbox[b]{0.66\columnwidth}{\flushleft%
(a) only real physical interactions are \hphantom{(a)}~considered;\\
(b) over-idealizations are not allowed.}
\end{equation}
It can safely be conjectured that for generic statistical operators $\rho$ and
Hamilton operators $H_1(t)$, the Humpty-Dumpty problem has no solution.
The examples of Sec.~\ref{sec:events} are to the point.

Maybe part of the evolutionary changes can be undone at least?
Consistent with rule (\ref{eq:0-60}a) we ask more specifically:
Can the two partial beams of a Stern--Gerlach apparatus
(SGA) be reunited with such precision that the initial spin state is
recovered? 
That is, we make no attempt at undoing in full the changes in the
center-of-mass state of the atom.
References \cite{Schwinger+2:88,Englert+2:88} deal with the details of such a
Stern--Gerlach interferometer; it is found that one needs three more SGAs for
the beam reunion.
Ideally, the four SGAs would be perfectly identical.
With due respect to rule (\ref{eq:0-60}b), however, we have to ask how large a
mismatch between the SGAs can be tolerated.

Two relevant quantities are the net transverse momentum transferred to either
partial beam and their net transverse displacement, respectively measured by
\begin{equation}\label{eq:0-61}
  \Delta p=\int\D t\,F(t) \AND \Delta z= -\int\D t\, F(t)\frac{t}{M}\,,
\end{equation}
where $F(t)$ is the force on the up-component, for instance, produced by the
inhomogeneity of the magnetic field.
Thus, $\Delta p$ and $\Delta z$ are properties of the macroscopic SGAs; the
deviations from the ideal values ${\Delta p=0}$, ${\Delta z=0}$ are resulting
from the lack of control over the SGAs.
It turns out that to maintain spin coherence one must, at least, ensure that
\begin{equation}\label{eq:0-62}
  \bigl(\delta z\,\Delta p/\hbar\bigr)^2
 +\bigl(\delta p\,\Delta z/\hbar\bigr)^2\ll1
\end{equation}
holds, where $\delta z$ and $\delta p$ are the natural spreads of the beam
prior to entering the first SGA.
Therefore, one can tolerate only such variations in the magnetic field that
both ${\olmagn{\Delta z}\ll\hbar/\delta p}$ and 
${\olmagn{\Delta p}\ll\hbar/\delta z}$, with the consequence
\begin{equation}\label{eq:0-63}
  \frac{\magn{\Delta z}\,\magn{\Delta p}}{\hbar}
  \ll\frac{\hbar}{\delta z\,\delta p}\,.
\end{equation}
This states that the \emph{macro}scopic magnetic field must be controlled with
\emph{sub-micro}scopic precision --- an impossible feat. 
The spin state is so severely affected by the inhomogeneous magnetic field of
the first SGA that it shares Humpty Dumpty's fate: One cannot undo the damage.

\section{State reduction}\label{sec:reduction}
Our statistical operator $\rho$ represents our knowledge
about the preparation of the system, and this knowledge does
not change until we acquire additional information.
When that happens, we must update the statistical
operator to account for the change in the conditions under which we predict
the results of future experiments.
\emph{State reduction} is the technical term for this updating.

State reduction is not particular to quantum theory, it is part and parcel of
all statistical formalisms.
Bookmakers know about state reduction, even if they never heard the term and
have not studied quantum theory.

Here is an example from the world of chess, showing two cases of state
reduction:%
\footnote{This is a slightly simplified account; for more details, see the news
  items on www.chessbase.com, 29-31 March 2013.}
After eleven rounds of the 2013 candidates tournament in London that would
determine the challenger of world champion Anand, a
prediction based on past results among the participants, their ratings, and
perhaps other data gave a chance of 87\% to Carlsen to win
the competition, a chance of 7\% to Kramnik, and a chance
of 6\% to Aronian.
In the twelfth round, Carlsen lost against Ivanchuk and
Kramnik won against Aronian, which resulted in updated chances of 65\% for
Kramnik and 35\% for Carlsen, with Aronian out of the race.
Then, in the thirteenth round, Carlsen won against
Radjabov and 
Kramnik drew his game with Gelfand, and the probabilities
for winning the tournament became 84\% for Carlsen and 16\% for Kramnik.
The highly improbable happened in the final fourteenth round:
Both Carlsen and Kramnik lost against Svidler and Ivanchuk,
respectively, and Carlsen took first place.

After the $k$th detector of a POM clicked, we update the statistical operator
in accordance with
\begin{equation}
  \label{eq:0-20}
  \rho\to\frac{K_k\phadj\rho K_k\adj}{p_k}\,,
\end{equation}
where $K_k$ is the appropriate \emph{Kraus operator} \cite{Kraus:83} for the
probability operator $\Pi_k$,
\begin{equation}
  \label{eq:0-21}
  \Pi_k=K_k\adj K_k\phadj\,.
\end{equation}
Depending on how the POM is implemented, there can be quite different Kraus
operators for the probability operator of interest.
In the example of Sec.~\ref{sec:MeasBorn}, we have
\begin{equation}
  \label{eq:0-22}
  K_1=\ket{\textsc{vac}}\,\sqrt{\eta_1/2\,}\,\bra{\pol{H}}
\end{equation}
for detector $\mathrm{D}_1$ and similar expressions for $\mathrm{D}_2$,
$\mathrm{D}_3$, and $\mathrm{D}_4$.
Irrespective of which detector clicks, we get
\begin{equation}
  \label{eq:0-23}
  \rho\to\ketbra{\textsc{vac}}{\textsc{vac}}
\end{equation}
upon reducing the state, thereby correctly stating that the photon has been
absorbed and we are left with the ``no more photon'' situation of the photon
vacuum, symbolized by the state ket $\ket{\textsc{vac}}$.

More complicated cases are possible, such as a sum of several $K\adj K$ terms
instead of the single term in \eqref{eq:0-21}.
The folklore that ``a measurement leaves the system in the relevant eigenstate
of the observable'' applies only to over-idealized projective measurements
(meaning that the $K_k$s are pairwise orthogonal projectors). 
It is puzzling that some textbook authors consider it good pedagogy to elevate
this folklore to an ``axiom'' of quantum theory.

``Collapse of the state'' or ``wave function collapse'' are popular synonyms
for state reduction.
The connotation that the transition in \eqref{eq:0-20} is a dramatic dynamical
process, as if the physical system were evolving, is clearly misleading.
The various mathematical models that, exactly or approximately, accomplish the
state reduction \eqref{eq:0-20} as a process in time result from a category
mistake: 
The statistical operator is not a physical object or a property of a physical
object, it \emph{describes} the object by encoding what we know about it, and
state reduction is the bookkeeping device for updating the description.
An electron carries its mass and charge around, but not ``its'' statistical
operator. 
It is worth recalling here that different physicists may very well use
different, equally correct, statistical operators for predictions about that
same electron.

\section{No action at a distance}\label{sec:notspooky}
Here is a quote from a recent news item:
\begin{equation}
\label{eq:0-Q1}
  \parbox{0.8\columnwidth}{%
``Quantum theory makes the distinctive prediction that non-local correlations 
are instant: for example, a measurement of the polarization of one of a pair 
of quantum-entangled photons should immediately set the polarization of the 
other, no matter how far the photons are apart, without either photon's 
polarization being in any way predetermined.''}
\end{equation}
In fact, quantum theory makes no such prediction.
There is no instant action at a distance.

\begin{figure*}[t]
\centerline{\includegraphics{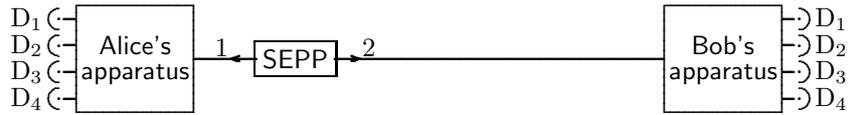}} 
\caption{\label{fig:EPRB-4POM}%
A source of entangled photon pairs (\textsf{SEPP}) sends photon~1 of each pair
to Alice's laboratory and photon~2 to Bob's laboratory. 
Alice's apparatus detects photon~1 before photon~2 reaches Bob's apparatus.
}
\end{figure*}

The situation alluded to is of the kind depicted in Fig.~\ref{fig:EPRB-4POM}.
A source of entangled photon pairs (SEPP) emits correlated photons in pairs,
with photon~1 sent to Alice and photon~2 to Bob who, for the sake of the
argument, is assumed to be much farther away from the SEPP than she is.
For the purpose of this discussion, we allow ourselves the idealization that
the photon pairs are described by a rank-one statistical operator,
specifically
\begin{equation}\label{eq:0-24}
\rho_{1\&2}^{\ }=\ket{\textsc{eprb}}\bra{\textsc{eprb}} 
\end{equation}
with an  Einstein--Podolsky--Rosen \cite{Einstein+2:35} state ket in Bohm's 
version \cite{Bohm:51}, 
\begin{equation}
  \label{eq:0-25}
  \ket{\textsc{eprb}}
=\frac{1}{\sqrt{2}}\bigl(\ket{\pol{HV}}-\ket{\pol{VH}}\bigr)
=\frac{1}{\sqrt{2}}\bigl(\ket{\pol{RL}}-\ket{\pol{LR}}\bigr)\,,
\end{equation}
where, for instance, the ket $\ket{\pol{HV}}$ stands for ``photon~1 is
$\pol{H}$-polarized, photon~2 is $\pol{V}$-polarized.''
Alice and Bob measure their respective photons by four-outcome POMs of the
construction shown in Fig.~\ref{fig:4POM}, with her detector clicking much
earlier than his.

Bob's statistical operator for photon~2 is that of a completely mixed
polarization state,
\begin{equation}
  \label{eq:0-26}
  \rho_2^{(\mathrm{B})}=\tr[1]{\rho_{1\&2}^{\ }}
=\half\bigl(\ketbra{\pol{H}}{\pol{H}}+\ketbra{\pol{V}}{\pol{V}}\bigr)
=\half\bigl(\ketbra{\pol{R}}{\pol{R}}+\ketbra{\pol{L}}{\pol{L}}\bigr)\,.
\end{equation}This is also Alice's statistical operator for photon~2 prior to
measuring photon~1.

Now suppose Alice's detector $\mathrm{D}_1$ clicks for photon~1.
Then, her reduced state for the photon pair is
\begin{equation}
  \label{eq:0-27}
  \rho^{(\mathrm{A})}=\ketbra{\textsc{vac}\;\pol{V}}{\textsc{vac}\;\pol{V}}\,,
\end{equation}
meaning that really only photon~2 is left, and Alice's updated statistical
operator for that photon is
\begin{equation}
  \label{eq:0-28}
   \rho_2^{(\mathrm{A})}=\tr[1]{\rho^{(\mathrm{A})}}=\ketbra{\pol{V}}{\pol{V}}\,.
\end{equation}
Has anything happened to photon~2 when Alice detected photon~1 with her
apparatus? 
Has ``the measurement \dots\ immediately set the polarization'' of photon~2?
Of course not.
The transition
\begin{equation}
  \label{eq:0-29}
  \half\bigl(\ketbra{\pol{H}}{\pol{H}}+\ketbra{\pol{V}}{\pol{V}}\bigr)
  \to\ketbra{\pol{V}}{\pol{V}}
\end{equation}
is Alice's bookkeeping, it is not a physical process that photon~2 is
undergoing, and it happens when she updates her statistical operator.

Incidentally, we have here an example of the situation mentioned twice above,
at the end of Sec.~\ref{sec:GenMeas} and at the end of Sec.~\ref{sec:reduction}:
Alice and Bob use different statistical operators for predictions about
photon~2, reflecting the different knowledge they have about the situation.
Alice knows the outcome of the measurement on photon~1, Bob does not.
And we must not fall into the trap of concluding that Alice's statistical
operator is better than Bob's;
both her $\rho_2^{(\mathrm{A})}$ and his $\rho_2^{(\mathrm{B})}$ give correct
statistical predictions, as can be verified by many repetitions of the
experiment.%
\footnote{A more complicated example is discussed in Sec.~4.3 of
  \cite{Englert+1:02}.} 

So, what about the quote \eqref{eq:0-Q1}?
It is another instance of the category mistake noted in
Sec.~\ref{sec:reduction}.
A change in Alice's description of photon~2 is misunderstood as a physical
process affecting photon~2, and then \foreign{ex falso sequitur quodlibet},
namely the false assertion about instantaneous nonlocal correlations,
exemplified by the immediate setting of the polarization of a photon that can
be arbitrarily far away.

\section{No nonlocality}\label{sec:local}
Nonlocal correlations are a familiar consequence of local actions:
A timing signal from a radio station synchronizes clocks at large distances
from each other.
These nonlocal correlations are clearly not instant;
they have a common local cause.
But then there are claims that quantum systems can exhibit correlations of
nonlocal origin. 

Recall, for instance, the abstract of a recent paper, which begins with these 
words:
\begin{equation}
\label{eq:0-Q2}
  \parbox{0.8\columnwidth}{%
``Bell's 1964 theorem, which states that the predictions of quantum theory
cannot be accounted for by any local theory, represents one of the most
profound developments in the foundations of physics.''}
\end{equation}
We submit: Doesn't quantum theory itself, which is a local theory, 
account for its own predictions?

As the authors of this quote know very well, experimental data contradict
Bell's theorem \cite{Bell:64,Clauser+3:69}, which implies that 
--- as a statement about physical systems --- the theorem is wrong.%
\footnote{We leave it as a moot point whether the theorem is ``one of the most
  profound developments in the foundations of physics.''} 
Since there is no error in the reasoning that establishes the theorem from its
assumptions, the flaw must be in the assumptions.
Specifically, it is the assumption that a mechanism exists that determines
which detector will click for the next photon registered by an apparatus of
the kind depicted in Fig.~\ref{fig:4POM}.
There is no such deterministic mechanism --- quantum processes are fundamentally
probabilistic, events are randomly realized ---
and the violation of Bell's theorem by actual data confirms that. 

Insights of this kind predate Bell's work by decades. 
Recall, for example, the matter-of-fact statement in 
Schr\"o\-din\-ger's essay of 1935, which reads
\begin{equation}
\label{eq:0-Q3}
  \parbox{0.8\columnwidth}{``[If I wish to ascribe] to \emph{all} determining
    parts definite (merely not exactly known to me) numerical values, then
    there is no supposition as to these numerical values \emph{to be imagined}
    that would not conflict with some portion of quantum theoretical
    assertions.'' }
\end{equation}
in Trimmer's translation \cite{Schrodinger:35}.

In a derivation of Bell's theorem, or any of the many
variants on record, the formalism of quantum theory is not used.
Instead, one relies on common-sense arguments of a certain plausibility 
in an attempt at describing quantitatively the properties of the joint
probabilities in experiments similar to that of Fig.~\ref{fig:EPRB-4POM}.
Part of that common sense is a certain notion of locality:
If the beam splitter in Fig.~\ref{fig:4POM} is replaced by a mirror that the
experimenter may or may not put in place, the free-will decisions by Alice
should have no influence on the click frequencies that Bob records in his
experiment, and \foreign{vice versa}.

The actual joint probabilities --- which are experimentally determined by
measuring many pairs emitted by the SEPP and are correctly accounted for by 
quantum theory ---  do not obey the restrictions that follow from that
common-sensible adhockery, and why should they?
Findings of an inadequate nonquantum formalism are irrelevant for quantum
physics.    
If the findings are at variance with the experimental data, as is the case
here, we are reminded of the inappropriateness of the reasoning.
It follows that common sense of that sort does not apply in the
quantum realm.

Rather disturbingly, though, it has become acceptable to turn the argument
into its opposite. 
It is taken for granted that quantum physics \emph{should} obey such common
sense, but then that inadequate nonquantum formalism needs nonlocal features
--- or so it seems.%
\footnote{One could also conclude that there is no deterministic mechanism,
  were this conclusion not prevented by that common sense.}
The conclusion that
\begin{equation}
\label{eq:0-Q4}
  \parbox{0.55\columnwidth}{``Quantum theory is nonlocal.''}
\end{equation}
appears inevitable:
So reads the opening line of a paper that reports a violation of
Bell's theorem by experimental data. 

In fact, however, there is no nonlocality in an experiment such as the one
sketched in Fig.~\ref{fig:EPRB-4POM} or any other quantum-physics experiment
on record.
The SEPP uses well-localized short-range interactions for the creation of the
photon pairs and the same is true for the processes by which the photon
absorption in one of the detectors gives rise to the observed click.
Quantum theory is a \emph{local theory}
in exactly the same sense as all other physical theories:
All interactions come about by local couplings; a continuity equation states
that probability is locally conserved; energy, momentum, angular momentum, and
other properties are transferred between particles by highly localized
scattering events, and so forth.
And just like the other local theories do, quantum theory predicts nonlocal
correlations that originate in local processes. 

It is true that joint probabilities that result from quantum processes can have
stronger correlations than those available by nonquantum simulations; 
a violation of Bell's theorem or one of its variants tells
us when such a situation is at hand.
This can be exploited for various purposes, including quantum key distribution
\cite{Ekert:91},
and experimental realizations \cite{Ling+5:08} have no nonlocal elements.
Accordingly, there are real-life practical applications of
Bell's theorem, and they give Bell's work lasting value. 

We close this subject with noting that a very basic notion of \textit{Local
  Quantum Physics} (the title of Haag's book \cite{Haag:96}) is precisely the
locality concept of Bell's theorem, namely that ``the
free-will decisions by Alice should have no influence on the click frequencies
that Bob records in his experiment, and \foreign{vice versa}.''
In the technical terms of relativistic quantum field theory, this is the
requirement that the local observables in Alice's space-time region commute
with the local observables in Bob's region, which is at a space-like
separation from Alice's.
Since the notion of locality that enters Bell's theorem is
exactly the notion of locality in quantum field theory, the observed violation
of Bell's theorem by quantum systems cannot support the
claim \eqref{eq:0-Q4}.

\section{Heisenberg's cut}\label{sec:H-cut}
\subsection{The cut is where you put it}\label{sSec:cut-location}
Alice and Bob have conducted the experiment of Fig.~\ref{fig:EPRB-4POM} on a
certain number of entangled photon pairs, independently and identically
prepared by the SEPP, and now they report their measurement results to Charlie.
What they communicate to him is the sequence of detector clicks observed and
the times at which the clicks occurred. 
These data enable Charlie to pair Alice's clicks with Bob's and so verify that
the click statistics observed is consistent with the predictions obtained from
applying Born's rule to this situation.

Alice and Bob will surely not trouble Charlie with accounts of their minds,
their retinas (or eardrums), or other irrelevant information.
Nor will Charlie bother with such matters when computing probabilities.
Yet, exactly this is what certain dogmatists insist must be done
\emph{in principle}, although \emph{in practice} they themselves never do it.%
\footnote{\label{fn:in-principle}%
  Statements like ``In principle, I could solve the 
  Schr\"odinger
  equation to predict the next solar eclipse.'' are empty unless you can do it
  in practice. Something can be \emph{impossible in principle}, such as
  drawing a triangular circle or building a so-called Popescu--Rohrlich
  box \cite{Popescu+1:94}, or \emph{possible in practice}.  
  But to declare that something is \emph{possible in principle} can at best
  mean that one is not aware of any reason that would make it impossible in
  principle.} 

The BS, the QWP, the PBSs of the apparatus in Fig.~\ref{fig:4POM} are made of
atoms with which the incoming photon interacts.
The detectors $\mathrm{D}_1$, \dots, $\mathrm{D}_4$ are made of atoms, too,
and so is the experimenter's retina and her brain.
If we apply quantum theory to the photon, we should also apply it to the atoms
of the optical elements, the detectors, \dots, the experimenter's brain.
And why stop here? 
There are also the atoms of your (the reader's of these lines) retina, optical
nerve, and brain. 
A truly complete account of the experiment should include all that --- as a
matter of principle, that is, not of practice.

No, it shouldn't. 
The experiment of Fig.~\ref{fig:EPRB-4POM} investigates the polarization
properties of entangled photon pairs, and the phenomena of interest do not
depend on which material is used to realize the BS and the other components of
the four-outcome POM of Fig.~\ref{fig:4POM}.
Yes, if one wishes to do so, one can extend the quantum description beyond the
crucial degrees of freedom of the photons, but nothing relevant will be
added to Charlie's analysis of the data.

In practice, the formalism of quantum theory is always applied to the relevant
quantum variables (almost always these are few) and the inclusion of
irrelevant ones increases the effort only but not the benefit.%
\footnote{This is also true for any application of Newton's classical
mechanics, Maxwell's electromagnetism, or any other physical
theory, without the dogmatists crying foul.}   
The identification of the relevant variables is usually unproblematic and, if
in doubt, one can take more quantum degrees of freedom into account than might
be necessary.
There are always plenty of technical details that are important for the
success of the experiment without entering the quantum-theoretical description
of the physical system that is studied.
For example, the experimenter has to calibrate the detectors and determine
their efficiencies, perhaps exploiting a mathematical model of some crucial
components, but in the end the only significant number is the efficiency
$\eta_k$ for the $k$th detector.
When Charlie analyzes the data, he must be able to trust the efficiencies
reported to him, but what kind of detector is actually used does not matter to
him.

The imagined demarcation line between the degrees of freedom that are given a
full quantum-theoretical description and the ``rest of the universe'' (to use
a pompous expression) is known as 
\emph{Heisenberg's cut},
and the inclusion of a few more variables into the quantum-theoretical
description than really necessary is the \emph{shift of Heisenberg's cut}. 

In a manner of speaking, one can regard Heisenberg's cut as
separating the microscopic realm of quantum physics from the macroscopic world
of classical physics, but one should not be carried away by the connotations
that the micro/macro distinction may bring about.
We are always referring to a particular context, such as Alice and Bob taking
data and Charlie analyzing them, not to a generally valid demarcation line.
When applying the quantum-theoretical formalism to an experiment in Munich,
the polarization degrees of freedom of photon pairs emitted by a SEPP in
Singapore are on the macroscopic side of Heisenberg's cut, whereas
they are on the microscopic side when the Singapore experiment is evaluated.

In his posthumous eloquent polemic \textit{Against `measurement'} 
\cite{Bell:90}%
\footnote{The replies by van Kampen \cite{vanKampen:90/91}, Peierls
  \cite{Peierls:91}, and Gottfried \cite{Gottfried:91} must not be missed.}
and also on earlier occasions,
Bell\footnote{This one exception from the no-finger-pointing policy of
  \fnref{NFP} is unavoidable.}
talks about Heisenberg's cut as if it were a universal border between quantum
physics and classical physics. 
Having made this mistake,%
\footnote{Bell also makes that category mistake of regarding
  the wave function as a physical object, and he misunderstands state
  reduction as a physical process.}
he then concludes that quantum theory is
ill-defined because it does not specify the location of the cut with the
mathematical precision that he demands it should have.

Bell is missing three essential points.
First, there is a Heisenberg's cut whenever the quantum-theoretical
formalism is applied in a particular situation; there is no such universal
border. 
Second, the cut is imprecise by its nature, it can be shifted to include as
many quantum degrees of freedom as one can handle in a calculation or control
in an experiment: 
Heisenberg's cut is where you put it.
Third, the distinction between microscopic phenomena and macroscopic phenomena
is meaningful even if one cannot give rigorous criteria for the micro/macro
distinction. 

We illustrate this remark by the following analogy.
The concepts of land and sea are meaningful although you cannot draw, in a
unique and natural way, a mathematical line on the beach that has the sea on
one side and the land on the other.
You can draw lines such that all of the sea is assuredly on one side, or all
of the land is on the other, but there is no line that accomplishes both.

A fourth point is this:
Alice and Bob rely on the macroscopic world around them, with its well-defined
properties, when they record their data and communicate them to Charlie.
One cannot even speak about quantum phenomena without reference to the rather
robust classical-physics environment, in which all human activity happens.
Denying that there is this robust environment, is denying the obvious.

In summary, then, Heisenberg's cut is needed and, by its nature, 
it is imprecise, and this is just fine.

\subsection{Decoherence}\label{sSec:decoherence}
Whenever we apply the formalism of quantum theory to a particular physical
situation, we identify the relevant degrees of freedom and ignore all the
rest.
In other words, we put Heisenberg's cut where we find it appropriate.
This involves a physical approximation because we disregard all dynamical
effects that arise from interactions between the quantum degrees of freedom
that are treated in full and those on the other side of the cut, often
referred to as the \emph{environment}.
As a consequence, the predictions we make about future measurements on the
system are not absolutely reliable and, since small errors accumulate, the
predictions are less reliable for the far future than the near future.

In terms of the parametric time dependence of the statistical operator, this
means that the von Neumann equation \eqref{eq:0-13} applies only
during the initial period when the residual interactions with the
environment can be ignored.
If we wish to extrapolate from the initial time $t_0$ to much later times, the
uncertainty that originates in our lack of knowledge about the
system-environment interactions across Heisenberg's cut has to be
taken into account.
This leads to a degradation of our statistical operator at later times~$t$,
usually noticeable as an increase of the entropy.
There is a rich literature about this so-called \emph{decoherence}, and many
ways of modeling the environment and the system-environment interactions have
been proposed and studied (a simple example of ``phase decoherence'' appears
in Sec.~\ref{sec:noproblem} below).
When a suitable model is adopted, the resulting modification of the 
von Neumann equation \eqref{eq:0-13} can give quite an accurate
account of the loss in precision of our predictions about the future
properties of the quantum system.

\section{No murky interpretation}\label{sec:notmurky}
The introduction of a recent paper contains the following sentences:
\begin{equation}
\label{eq:0-Q5}
  \parbox{0.8\columnwidth}{``[\dots] a century after the discovery
    of quantum mechanics, it seems that we are no closer to a consensus about
    its interpretation than we were in the beginning. The collapse of the
    quantum state occurring during the process of measurement [\dots]
    does not have an unambiguous definition and a reasonable
    explanation. 
    [\dots] some radical changes in our classical understanding of reality
    have to be made; e.g. constructing a physical process of collapse,
    accepting the existence of parallel worlds, or adding nonlocal hidden
    variables.''} 
\end{equation}
Of course, the authors of these lines use Born's rule for the
calculation of probabilities, associate the wavelengths of spectral lines with
differences of energy eigenvalues, and use all other standard links between
the mathematical formalism and the physical phenomena.
In other words, these authors' interpretation of quantum theory is
the usual one.%
\footnote{\label{fn:THE}%
It is simply \emph{the} interpretation of quantum theory. 
Calling it the standard interpretation or using any other identifier is
unnecessary and, indeed, harmful because ``this sounds as if there were
several interpretations of quantum mechanics.
There is only one.'' 
(Peierls as quoted by Davies and Brown \cite{Davies+1:86})} 
Why, then, is there this blunt assertion that ``there is no consensus about the
interpretation''?  

Clearly, the authors of \eqref{eq:0-Q5} have a nonstandard meaning of the word
``interpretation'' in mind, and so have others.%
\footnote{The \textit{Compendium of Quantum Physics}
  \cite{Greenberger+2.ed:2009} has entries on more than ten
  ``interpretations,'' and two recent polls \cite{Schlosshauer+2:13,Sommer:13}
  gave about that many ``interpretations'' to choose from, while 
  Tegmark's poll of 1998 \cite{Tegmark:98} had only half as many choices.} 
The actual interpretation, that is: the battle-tested link between the
formalism and the phenomena, is not what they are questioning.
They would refer to that as the ``minimal interpretation''
and that is not enough for them.
They want to know ``what is really going on,'' thereby requiring answers to
questions that quantum theory cannot provide.

These are questions about the \emph{mind-body problem}, the relation between
our conscious human experiences and our free will on one side, and the reality
of natural phenomena on the other ---
philosophical issues that are outside of physics.
Yes, it is important to understand the philosophical implications of the
lessons of quantum physics, but one must not confuse physics with philosophy.

The ``lack of consensus'' noted in \eqref{eq:0-Q5} is not about the
interpretation of quantum theory in the strict sense of Sec.~\ref{sec:PFI},
but about its bearing on philosophy.
The use of the term ``interpretation'' for the more or less
successful reconciliation of philosophical preconceptions with the
quantum-physical phenomena is most unfortunate, and there is little hope that
this practice will change.
Fuchs's and Peres's verdict \textit{Quantum Theory Needs No `Interpretation'}
(the title of their essay \cite{Fuchs+1:00}) 
should be understood in this context: 
Quantum theory has its interpretation and does not need a philosophical 
\foreign{\"Uberbau}. 

Of course, philosophers should be encouraged to study the lessons of quantum
theory --- for example, the lesson that the future is not predetermined by the
past\footnote{\label{fn:CannotKnow}%
  However the next photon is prepared, there is no way of knowing
  which of the four detectors in the experiment of Fig.~\ref{fig:4POM} will
  click.} --- but their philosophical debates
are really irrelevant for quantum theory as a physical theory, although they
are likely to be of great interest to physicists. 
We could leave it at that, were there not the widespread
habit of the debaters to endow the mathematical symbols of the formalism
with more meaning than they have.
In particular, there is a shared desire to regard the 
Schr\"odinger wave
function as a physical object itself after forgetting, or refusing to accept,
that it is merely a mathematical tool that we use for a description of the
physical object (electron, atom, photon, \dots).
The regrettable textbook practice of invoking analogies between
Schr\"odinger's
probability-amplitude waves and sound waves, water-surface
waves, or electromagnetic waves provokes and feeds that desire.
The fundamental difference is then ignored:
Sound waves, water waves, electromagnetic waves are physical objects with
mechanical properties, they carry energy, momentum, angular momentum from here
to there, whereas the Schr\"odinger
waves are mathematical symbols.%
\footnote{Also, only very special cases of wave functions are
  probability-amplitude waves in the three-dimensional space of physical
  positions.} 

The reification of the wave function requires that the ``collapse of the
quantum state'' is a physical process, but since it isn't, 
there cannot be ``an unambiguous definition,'' and a
``reasonable explanation'' is not possible.
We submit:
With such incorrect meaning given to the wave function and to the process of
state reduction, one no longer deals with quantum theory, but with something
else. 
The conclusions that are then reached about philosophical implications are of
no consequence.

Indeed, the alleged implications (``parallel worlds'') are bewildering and
prompt us to paraphrase van Kampen's Theorem IV \cite{vanKampen:88}:
If you endow the mathematical symbols with more meaning than they have,
you yourself are responsible for the consequences, and you must not blame
quantum theory when you get into dire straits.

\emph{The} interpretation of quantum theory is not murky, 
it is absolutely clear;
there is no lack of consensus about it.
It is one of the three defining constituents of quantum theory
(Sec.~\ref{sec:PFI}), and all practicing quantum physicists rely on it all the
time.

\section{You cannot see what you wouldn't recognize}
\subsection{No measurement problem}\label{sec:noproblem}
We can use the set-up of Fig.~\ref{fig:4POM}, 
see Secs.~\ref{sec:MeasBorn-1} and \ref{sec:MeasBorn-2}, for a discussion of
the so-called ``measurement problem.''
A typical argument would develop along the following lines.
At the initial time $t_{\mathrm{ini}}$, we have the photon approaching the BS
with a polarization-state ket
\begin{equation}\label{eq:0-30}
  \ket{\wp}=\ket{\pol{V}}\alpha+\ket{\pol{H}}\beta \WITH
\magn{\alpha}^2+\magn{\beta}^2=1\,,
\end{equation}
and the four detectors are in the ``ready'' state
\begin{equation}\label{eq:0-31}
  \ket{\Dstate{R}}=\ket{\Dstate{R}_1\Dstate{R}_2\Dstate{R}_3\Dstate{R}_4}\,,
\end{equation}
so that we write
\begin{equation}\label{eq:0-32}
  \ket{\ }=\ket{\wp\,\Dstate{R},t_{\mathrm{ini}}}
\end{equation}
for the state ket as characterized by properties referring to the initial
time.
Expressed with reference to the final time $t_{\mathrm{fin}}$, this state ket
is
\begin{equation}\label{eq:0-33}
  \ket{\ }=\ket{\textsc{vac}\;\Dstate{NC},t_{\mathrm{fin}}}\sqrt{p_0}
+\sum_{k=1}^4\ket{\textsc{vac}\;\Dstate{C}_k,t_{\mathrm{fin}}}\sqrt{p_k}\,,
\end{equation}
where $\Dstate{C}_k$ stands for ``the $k$th detector has clicked and
the other three have not'' and $\Dstate{NC}$ denotes the no-click
situation. 
At the final time, the photon has been absorbed and we have the photon vacuum
state for all detector states, which allows us to focus on the final state of
the detectors,
\begin{equation}
  \label{eq:0-34}
  \ket{\ }=\ket{\textsc{vac}\;\Dstate{D},t_{\mathrm{fin}}} 
\end{equation}
with
\begin{equation}
  \label{eq:0-35}
  \ket{\Dstate{D}}=\ket{\Dstate{NC}}\sqrt{p_0}
           +\sum_{k=1}^4\ket{\Dstate{C}_k}\sqrt{p_k}\,.
\end{equation}
This is the point where some start to worry: 
The detectors are in a superposition of this one having clicked or that one
having clicked, then how does one account for a definite click of one of the
detectors and why do we never see superpositions such as 
$\bigl(\ket{\Dstate{C}_1}+\ket{\Dstate{C}_2}\bigr)/\sqrt{2}$
with the two detectors having half-clicked?

Although these questions seem to be addressing profound issues, they are not
formulating a serious ``measurement problem.''
At best, they are rephrasing matters already discussed.

\textbf{First}, we note that the state ket $\ket{\ }$ in \eqref{eq:0-34} is
the same as the one in \eqref{eq:0-32}.
The two versions of $\ket{\ }$ are particular examples of the general
statement in \eqref{eq:0-18b}. 
Both versions of $\ket{\ }$ encode what we know about the physical situation.
Why, then, are all those questions asked about \eqref{eq:0-34}, but none about
\eqref{eq:0-32}? 

\textbf{Second}, the superposition 
$\bigl(\ket{\Dstate{C}_1}+\ket{\Dstate{C}_2}\bigr)/\sqrt{2}$
does not describe a situation where detectors $\mathrm{D}_1$ and
$\mathrm{D}_2$ each have ``half-clicked'' (whatever that would mean) but an
entirely new situation that is outside our experience.
Remember that $\bigl(\ket{\pol{V}}+\ket{\pol{H}}\I\bigr)/\sqrt{2}$ is not a
polarization state that is ``half-vertical'' and ``half-horizontal'' but one
that is entirely right-circular, phenomenologically very different from both
$\ket{\pol{V}}$ and $\ket{\pol{H}}$.
We know what  $\bigl(\ket{\pol{V}}+\ket{\pol{H}}\Exp{\I\varphi}\bigr)/\sqrt{2}$
means (as a statement about a photon's polarization) for any value of the
phase $\varphi$, but we have no clue about 
$\bigl(\ket{\Dstate{C}_1}+\ket{\Dstate{C}_2}%
\Exp{\I\varphi}\bigr)/\sqrt{2}$.
So, we never see these superpositions because we cannot recognize them.%
\footnote{\label{fn:heWho1}%
  He who interjects ``But, in principle, I could certainly measure the
  observables described by the hermitian operators
  ${\ket{\Dstate{C}_1}\Exp{\I\varphi}\bra{\Dstate{C}_2}
  +\ket{\Dstate{C}_2}\Exp{-\I\varphi}\bra{\Dstate{C}_1}}$,
  and when the eigenvalue $+1$ is found, I will have succeeded.'' should
  reread \fnref{in-principle} and then ask himself 
  \emph{Can I do it in practice?}} 
If we could, either version of $\ket{\ }$ would enable us to calculate the
probability of seeing these superpositions.

\textbf{Third}, despite strong assertions that it does,
decoherence does not come to the rescue.
The asserters consider the statistical operator associated with ket
$\ket{\Dstate{D}}$, 
\begin{eqnarray}\label{eq:0-36}
  \rho^{\ }_{\mathrm{D}}&=&\ket{\Dstate{D}}\bra{\Dstate{D}}
\\\nonumber &=&\ket{\Dstate{NC}}p_0\bra{\Dstate{NC}}
+\sum_{k=1}^4\ket{\Dstate{C}_k}p_k\bra{\Dstate{C}_k}
+\bigl\{\textrm{cross terms}\bigr\}\,,
\end{eqnarray}
where the cross terms collect all ket-bra products of two different outcomes,
\begin{eqnarray}\label{eq:0-37}
  \bigl\{\textrm{cross terms}\bigr\}&=&\sum_{k=1}^4
\bigl(\ket{\Dstate{NC}}\sqrt{p_0p_k}\bra{\Dstate{C}_k}
+\ket{\Dstate{C}_k}\sqrt{p_0p_k}\bra{\Dstate{NC}}\bigr)
\nonumber\\\mbox{}&&
+\sum_{j\neq k}\ket{\Dstate{C}_j}\sqrt{p_jp_k}\bra{\Dstate{C}_k}\,.
\end{eqnarray}
At this point, one remembers that $\ket{\Dstate{C}_k}$ symbolizes a
state of very many quantum degrees of freedom (all those nuclei and electrons
that make up the optical elements and the detectors) 
and, therefore, one is suddenly unsure about details such as the
over-all phase assigned to $\ket{\Dstate{C}_k}$, and so it appears
natural to refine the quantum-theoretical description by the replacements 
$\ket{\Dstate{C}_k}\to\ket{\Dstate{C}_k}\Exp{\I\phi_k}$
with random phases $\phi_k$.
Here, ``random'' simply means that the phases acquire different unknown values
whenever the experiment is repeated.
Averaging over the random phases
removes the cross terms,
\begin{equation}\label{eq:0-38}
  \bigl\{\textrm{cross terms}\bigr\}\to0\,,
\end{equation}
and we arrive at an effective statistical operator for the detectors,
\begin{equation}\label{eq:0-39}
   \rho^{\ }_{\mathrm{D}}\to\rho^{\ }_{\mathrm{D, eff}}
=\ket{\Dstate{NC}}p_0\bra{\Dstate{NC}}
+\sum_{k=1}^4\ket{\Dstate{C}_k}p_k\bra{\Dstate{C}_k}\,.
\end{equation}
The adjective ``effective'' means that all observable properties calculated
with $\rho^{\ }_{\mathrm{D, eff}}$ are the same as
those calculated with $\rho^{\ }_{\mathrm{D}}$.
In other words, the cross terms of \eqref{eq:0-36} are ineffective, they are
of no consequence.
This, of course, is not a recent insight.
There are, for example, the accounts by S\"ussmann \cite{Sussmann:58} and Peres
\cite{Peres:80} of 1958 and 1980, respectively.

So, $\rho^{\ }_{\mathrm{D}}$ ``has decohered'' into $\rho^{\ }_{\mathrm{D,eff}}$ 
--- the transition \eqref{eq:0-39} is now regarded as a physical process! --- 
and we have a mixed state that is as if we had a 
Gibbs ensemble
composed of a fraction $p_0$ of no-click cases, a fraction $p_1$ of clicks by
detector $\mathrm{D}_1$, \dots, a fraction $p_4$ of clicks by detector
$\mathrm{D}_4$. 
One of the situations is the case, and we do not know which one until we take
note of the outcome of the experiment.
All is fine, or so it seems, because we just have this ``classical ignorance''
rather than the quantum-theoretical indefiniteness of  
$\rho^{\ }_{\mathrm{D}}$, and we no longer need to worry why we do not see
the detectors with properties described by superpositions such as  
$\bigl(\ket{\Dstate{C}_1}+\ket{\Dstate{C}_2}\bigr)/\sqrt{2}$.

All is fine? No, we are fooling ourselves.
Remembering Sec.~\ref{sec:as-if}, we note that 
the effective statistical operator does not have this one unique 
as-if reality associated with it, but very many of them.
For example, we have
\begin{eqnarray}\label{eq:0-40}
  \rho^{\ }_{\mathrm{D,eff}}&=&\ket{\Dstate{NC}}p_0\bra{\Dstate{NC}}
                        +\ket{\Dstate{C}_1}p_1\bra{\Dstate{C}_1}
                        +\ket{\Dstate{C}_2}p_2\bra{\Dstate{C}_2}
\nonumber\\&&\mbox{}
+\ket{\Dstate{C}_{34+}}\frac{p_3+p_4}{2}\bra{\Dstate{C}_{34+}}
\nonumber\\&&\mbox{}
+\ket{\Dstate{C}_{34-}}\frac{p_3+p_4}{2}\bra{\Dstate{C}_{34-}}\qquad
\end{eqnarray}
with
\begin{equation}\label{eq:0-41}
  \ket{\Dstate{C}_{34\pm}}=\ket{\Dstate{C}_3}\sqrt{\frac{p_3}{p_3+p_4}}
             \pm\ket{\Dstate{C}_4}\sqrt{\frac{p_4}{p_3+p_4}}
\end{equation}
and, clearly, the mixture described by $\rho^{\ }_{\mathrm{D,eff}}$ can also
be blended by $p_0$ parts of no-click cases, 
$p_1$ parts of clicks by $\mathrm{D}_1$,
$p_2$ parts of clicks by $\mathrm{D}_2$, plus $\half(p_3+p_4)$ parts each for
the situations (of unknown phenomenology) that are described 
by the superpositions $\ket{\Dstate{C}_{34+}}$ and $\ket{\Dstate{C}_{34-}}$.
As always, there is an infinity of blends for the one mixture and a
corresponding infinity of as-if realities.

Another reason why we are fooling ourselves is that we still need the
principle of random realization for selecting one of the five possibilities
and Born's rule for calculating the probabilities from 
$\rho^{\ }_{\mathrm{D,eff}}$.
It is not the case, as self-suggesting this may appear, that we can just read
off the probabilities by looking at the weights of our preferred as-if
reality. 
Such a reading-off can sometimes give the right answer, but we know that it
does only after an application of Born's rule has confirmed
our educated guess. 

\textbf{Fourth}, the transition \eqref{eq:0-39} is an improvement of our
description, it is not a physical process.
The decoherence argument with the unknown phase factors
$\Exp{\I\phi_k}$ and so forth is a confession that we don't know the dynamics
with sufficient precision, and our starting point
\eqref{eq:0-32} is wrong, too.
We do not have enough information for describing the initial ``ready''
state of the detectors by a state ket $\ket{\Dstate{R}}$ or its corresponding
rank-one statistical operator $\ket{\Dstate{R}}\bra{\Dstate{R}}$.
We can, at most, have limited knowledge about a few degrees of freedom of the
detectors and account for that by a highly mixed state.
The state kets of \eqref{eq:0-32}, \eqref{eq:0-34}, and
\eqref{eq:0-35} border on utter nonsense.

When preparing the experiment and getting the detectors into the ``ready''
state, we are not controlling all quantum degrees of freedom of the detectors
perfectly but, at best, the few relevant ones with some limited precision.
There are, therefore, very many pure quantum states that we could imagine for
the ``ready'' detectors rather than a unique state associated with the 
ket~$\ket{\Dstate{R}}$.
Strictly speaking, there is not even the pure-polarization ket~$\ket{\wp}$, 
but that is acceptable as a pretty good approximation in a
well-controlled experiment.

Rather than the convex sum of projectors in \eqref{eq:0-39} and
\eqref{eq:0-40}, we have, at best, a convex sum of highly mixed statistical
operators for the detectors,
\begin{equation}\label{eq:0-42}
  \rho^{\ }_{\mathrm{D}}=p_0\rho^{(\mathrm{NC})}_{\mathrm{D}}
                    +\sum_{k=1}^4p_k\rho^{(k)}_{\mathrm{D}}\,,
\end{equation}
where $\rho^{(k)}_{\mathrm{D}}$ is for a click of the $k$th detector and
$\rho^{(\mathrm{NC})}_{\mathrm{D}}$ for the no-click case.
But this $\rho^{\ }_{\mathrm{D}}$, too, is more an illusion than a reality.
Even the smallest detectors contain a gargantuan number of atoms, and
we don't really know what exactly the detectors are composed of ---
removing a single molecule, say, from the surface of one of the wires in the
apparatus has no bearing on the relevant aspects of the physical situation.
Therefore, we will be hard-pressed to write down meaningful expressions
for the $\rho^{(k)}_{\mathrm{D}}$s.%
\footnote{\label{fn:heWho2}He who is still pondering the question in footnote
\ref{fn:heWho1}, may find a useful hint here.}

This is not a problem, though, because there is no use of the quantum
description of the detectors that have clicked or not. 
There are no probabilities of subsequent events to be calculated from
$\rho^{\ }_{\mathrm{D}}$.
The photon-absorption event has been amplified or not, 
one of the detectors has clicked or the photon has escaped detection:
The measurement of this photon's polarization is over. 

\textbf{Fifth}, since neither decoherence nor any other mechanism select one
particular outcome (see Sec.~\ref{sec:local}), 
the whole ``measurement problem'' reduces to the question 
\emph{Why is there one specific outcome?} which is asking 
\emph{Why are there randomly realized events?} in the particular context
considered. 
This harkens back to Sec.~\ref{sec:PFI}, where we noted that quantum theory
cannot give an answer.

In summary, then, the alleged ``measurement problem'' does not exist as a
problem of quantum theory.
Those who want to pursue the question \emph{Why are there events?} must seek
the answer elsewhere.

\subsection{Schr\"odinger's cat}\label{sec:S-cat}
In this context one can hardly avoid the mentioning of Schr\"odinger's
infamous cat.%
\footnote{Heisenberg's dog \cite{Bergou+1:98} does not
  concern us here.}   
When commenting on the physical significance of the quantum-theoretical 
wave function --- it is an
\emph{expectation catalog} (\foreign{Katalog der Erwartung}) 
in Schr\"odinger's 
own words --- he gave, as a warning, 
the ``ridiculous'' (\foreign{burlesk}) example of a cat
whose vital state serves as the indicator whether a radioactive decay has
occurred or not.
That passage reads
\begin{equation}
\label{eq:0-Q6}
  \parbox{0.8\columnwidth}{``One can even set up quite ridiculous cases. 
   A cat is penned up in a steel chamber, along with the following diabolical 
   device (which must be secured against direct interference by the cat): 
   in a Geiger 
   counter there is a tiny bit of radioactive substance, \emph{so}
   small, that \emph{perhaps} in the course of one hour one of the atoms
   decays, but also, with equal probability, perhaps none; if it happens, the
   counter tube discharges and through a relay releases a hammer which
   shatters a small flask of hydrocyanic acid. If one has left this entire
   system to itself for an hour, one would say that the cat still lives
   \emph{if} meanwhile no atom has decayed. The first atomic decay would have
   poisoned it. The $\psi$-function of the entire system would express this by
   having in it the living and the dead cat (pardon the expression) mixed or
   smeared out in equal parts.''}
\end{equation}
in Trimmer's translation \cite{Schrodinger:35}.
In other words, the diabolic device establishes the strong correlations
\begin{eqnarray}\label{eq:0-43}
\textrm{no decay} &\longleftrightarrow& \textrm{live cat}\,,\nonumber\\
\textrm{yes decay} &\longleftrightarrow& \textrm{dead cat}\,,
\end{eqnarray}
so that the cat is a \foreign{bona fide} detector for the radioactive decay 
event.
It has the obvious advantage that the detector is easy to read.

Schr\"odinger is, of course,
overshooting the mark.
Just like there is no ket $\ket{\Dstate{D}}$ for the detectors in
Sec.~\ref{sec:noproblem}, there is no $\psi$-function that has ``the living
and dead cat smeared out.''
The imagined superpositions of the dead and the live cat, 
which have inspired so many authors, are fantastic phantoms of the same
kind as the detector states associated with the kets
$\bigl(\ket{\Dstate{C}_1}+\ket{\Dstate{C}_2}%
\Exp{\I\varphi}\bigr)/\sqrt{2}$.
We do not know how to recognize them and, therefore, we cannot see them. 
In all of that, the cat is really superfluous, it just adds drama to the plot; 
the position of the hammer, say, already tells the story.  

It would be fitting if any such correlated state of a quantum degree of
freedom and a macroscopic detector would be called a
\emph{Schr\"odinger-cat state}
in today's jargon.
But that is not the case.
The term is routinely applied to superpositions of
states of a few quantum degrees of freedom, or even of a single one, 
that have vastly different quantum numbers.
In this sense, the silver atom emerging from a 
Stern--Gerlach apparatus is in a  Schr\"odinger-cat state.
This usage of the term does not do justice to what 
Schr\"odinger had in mind;
it should be discouraged, the more so because such superpositions are
nothing special, they are common in standard interferometric devices.

\section{Summary}
The foundations of quantum theory have been laid, and laid well, 
by the founding fathers --- 
Planck, Einstein, Bohr, Heisenberg, Schr\"odinger, Born, Dirac, and others. 
Much has happened since those early days of the 1920s.
The phenomenology and the formalism are much richer today and, to supplement
this enrichment, the interpretation has more links between the formalism and the
phenomena.
This is a natural consequence of our communal duty to explore the
consequences, implications, and applications of quantum theory, and we are
still in the process of doing just that. 

It is true that, occasionally, the founding fathers and their peers found it
difficult to reconcile the lessons of quantum theory with the world view they
had acquired earlier. 
The younger players, notably Heisenberg and Dirac, 
overcame these difficulties more quickly and more easily than their seniors, 
Planck and Einstein among them.  
Einstein, for example, did not like the random nature of events.
He told Franck the following:%
\footnote{As quoted by Snow in \cite{French.ed:79}.} 
\begin{equation}
\label{eq:0-Q7}
  \parbox{0.8\columnwidth}{  ``I can, if the worse comes to worst, still realize that God may have created
  a world in which there are no natural laws. In short, chaos. But that there
  should be statistical laws with definite solutions, i.e., laws that compel
  God to throw dice in each individual case, I find highly disagreeable.''} 
\end{equation}
While Einstein's metaphor of the dice-throwing God is charming and
memorable, and his personal difficulties are of great historical interest, 
it is ultimately irrelevant what he, or any other individual, finds 
disagreeable.
The facts count --- and Einstein, of course, understood this.
In science, truth is the daughter of time, not of authority.

One must also acknowledge that, after the lapse of fourscore years, the
terminology is more precise today than it was during the founding years.
Concepts get clearer in time, and one learns to avoid sloppy terms and
misleading phrases.
The foundations, however, have not been touched by these refinements.

As way of summary, here are our answers to the questions asked at the
beginning:  
\begin{quotation}
\QandA{Yes, quantum theory well defined.}
\QandA{Yes, quantum theory has a clear interpretation.}
\QandA{Yes, quantum theory is a local theory.}
\QandA{No, quantum evolution is not reversible.}
\QandA{No, wave functions do not collapse; you reduce your state.%
\rule[-3pt]{0pt}{4pt}}
\QandA{No, there is no instant action at a distance.}
\QandA{Heisenberg's cut is where you put it.}
\QandA{No, Schr\"odinger's cat is not half dead and half alive.%
\rule[-3pt]{0pt}{4pt}}
\QandA{No, there is no ``measurement problem.''}
\end{quotation}
Tersely:
Quantum theory is a well-defined local theory with a clear interpretation. 
No ``measurement problem'' or any other foundational matters are waiting to be
settled. 

What, then, about the steady stream of publications that offer solutions for
alleged fundamental problems, each of them wrongly identified on the basis of
one misunderstanding of quantum theory or another?  
Well, one could be annoyed by that and join van Kampen \cite{vanKampen:08} in
calling it a scandal when a respectable journal prints yet another such
article.
No-one, however, is advocating censorship, even of the mildest kind, because
the scientific debate cannot tolerate it. 
Yet, is it not saddening that so much of diligent effort is wasted on
studying pseudo-problems? 

\bigskip\bigskip

\begin{small}\noindent%
The Centre for Quantum Technologies is a Research Centre of Excellence funded
by the Ministry of Education and the National Research Foundation of Singapore.
\end{small}


\newcommand{\arxiv}[3][quant-ph]{\textrm{eprint\ arXiv:#2[#1] (#3)}}

\enlargethispage{-7.8\baselineskip}

\end{document}